\documentclass[english,[english,12pt,prd,aps,final,notitlepage,nobibnotes,nofootinbib,superscriptaddress,showpacs]{revtex4-1}
\usepackage[T1]{fontenc}
\usepackage[latin9]{inputenc}
\usepackage{verbatim}
\usepackage{multirow}
\usepackage{amstext}
\usepackage{graphicx}

\makeatletter

\providecommand{\tabularnewline}{\\}

\usepackage[hidelinks]{hyperref}

\newcommand{\R}{\mathcal{R}}
\newcommand{\M}{\mathcal{M}}

\makeatother

\usepackage{babel}
\begin{document}
\title{Multiple Charged Meson Production in Exclusive $B_{c}$ Decays: $K+4\pi,$$KK+3\pi$,
$7\pi$ Cases.}
\author{A.V. Luchinsky}
\affiliation{Institute for High Energy Physics, Protvino, Russia}
\begin{abstract}
Theoretical analysis of three exclusive decays, $B_{c}\to\psi^{(')}+\R,$where
$R=K+4\pi,$$KK+3\pi$, or $7\pi$,  is given. Using the factorization
method and the resonance approximation (methods proved to be very
useful in the analysis of some other similar decays), we have obtained
analytical expressions of the required amplitudes and created distributions
over some interesting mass combinations. Presented results could be
used for comparison with forthcoming experimental data and better
understanding of the nature of heavy quarkonia.
\end{abstract}
\maketitle

\section{Introduction}

Heavy quarkonia mesons, that is particles that in valence approximation
are built from heavy (i.e. $c$ of $b$) pairs can be considered as
a unique laboratory to study strong interaction both in perturbative
and non-perturbative regimes. On the partonic level the processes
of their production and decay can be described in terms of creation
or annihilation of the heavy quark, that hadronizes later into experimentally
obesrved particle. Doubly heavy meson with an opened flavor,$B_{c}=bc$
take an intermediate position between charmonium $\left(c\bar{c}\right)$
and bottomonium $\left(b\bar{b}\right)$ particles, that allows one
to use it as a test for models that were developed to describe heacy
quarkonium particles with hidden flavor. Nice theoretical review concerning
different properties of this particle can be found, for example, in
paper \cite{Gershtein:1994jw}

The presented paper is devoted to phenomenological analysis of some
exclusive $B_{c}$-meson decay with the vector charmonium production
accomplished by $K+4\pi$, $KK+3\pi$ or $7\pi$. This work is a continuation
of series of papers at were devoted to production of other systems
of light mesons in the similar processes.

General idea used in all these works is simple. It is well known that
in valence approximation $B_{c}$-meson is built from $b$- and $c$-quarks.
Exclusive decay of this particle into vector charmonium and some set
of light mesons on the partonic level can be described by the weak
decay of the constituent $b$-quark: $b\to Wc$. Produced $W$ boson
hadronizes into the final system of light particles, while $c$-quark
with a spectator $\bar{c}$ forms $J/\psi$ or $\psi(2S)$ . The subprocess
$B_{c}\to W\psi^{(')}$ can be described in terms of weak $B_{c}$-meson
form-factors. The suitable theoretical model for the W- boson hadronization,
on the other hand, is the so-called resonance approach.

Described above simple model turned out to be surprisingly powerful.
It was used to describe decays $B_{c}\to J/\psi+3\pi$, etc., and
the predictions for differential distributions and integrated branching
fractions are in good agreement withe experimental results. In the
present paper we are extending the analysis to three new decays.

The rest of the paper is organized as follows. In the next section
a short description of the adopted approach is given and the parametrization
of the $B_{c}$-meson's form-factors is resented. Section 3 is devoted
to theoretical description of same observed already decays, that mile
be used as subprocesses for new ones. These new decays will be studied
in section 4, while the last section is reserved to some discussion
and conclusion.

\section{Exclusive $B_{c}$ decays}

\begin{figure}
\begin{centering}
\includegraphics[scale=0.5]{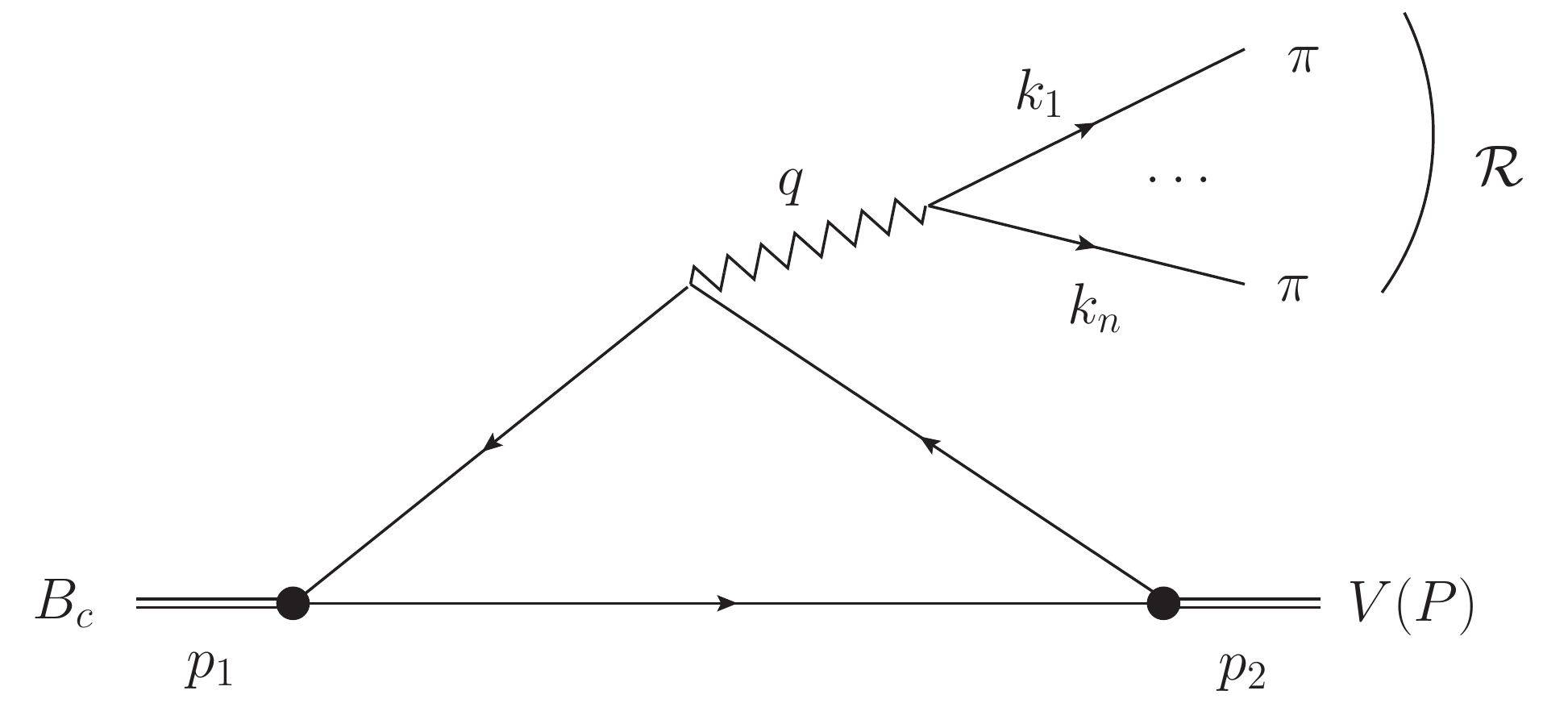}
\par\end{centering}
\caption{Feynman diagram for $B_{c}\to\psi\R$ decay\label{diag:BcJR}}
\end{figure}

In the partonic approximation $B_{c}$ meson is build from $b\bar{c}$
quarks, so its exclusive decays into vector charmonium and a set of
light mesons can be described as a weak decay of the constituent $b$
quark. Produced $c$ with a spectator $\bar{c}$ forms the final charmonium,
while $W$-boson hadronizes into the final system of light mesons
$\R$. The schematic Feynman diagram of this process is shown in figure
\ref{diag:BcJR}.

In the factorization approximation this diagram can be written in
the form
\begin{eqnarray*}
\M & = & \frac{G_{F}V_{bc}}{\sqrt{2}}a_{1}H^{\mu}\epsilon_{\mu}^{(\R)},
\end{eqnarray*}
where the Wilson coefficient $a_{1}(m_{b})\approx1.14$ describes
the effect of final state interaction \cite{{Buchalla:1995vs}}, $G_{F}$
is a Fermi coupling constant, $V_{bc}$ is a corresponding coefficient
of the CKM matrix, while $H^{\mu}$ and $\epsilon_{\mu}^{(\R)}$ are
the amplitudes of the $B_{c}\to\psi W$ and $W\to R$ transitions
respectively.

Let us define the the first amplitude first. It is clear, that it
can depend only on the momenta of the initial and final heavy mesons
$p_{1,2}$. There are several ways how it can be written in the Lorentz-invariant
form and in our paper we will use the parametrization adopted, for
example, in paper \cite{Kiselev:2000pp}:
\begin{eqnarray*}
H_{\mu} & = & \left[2M_{\psi}A_{0}\left(q^{2}\right)\frac{q^{\mu}q^{\nu}}{q^{2}}+\left(M_{B_{c}}-M_{\psi}\right)A_{1}\left(q^{2}\right)\left(g^{\mu\nu}-\frac{q^{\mu}q^{\nu}}{q^{2}}\right)-\right.\\
 &  & \left.A_{2}\left(q^{2}\right)\left(p_{1}^{\mu}+p_{2}^{\mu}-\frac{M_{B_{c}}^{2}-M_{\psi}^{2}}{q^{2}}q^{\mu}\right)-\frac{2iV\left(q^{2}\right)}{M_{B_{c}+M_{\psi}}}e^{\mu\nu\alpha}p_{1\alpha}p_{2}\right]\epsilon_{\nu}^{(\psi)}.
\end{eqnarray*}
Here $\epsilon^{(\psi)}$ is the polarization vector of final charmonium
meson and $A_{0,1,2}\left(q^{2}\right)$, $V\left(q^{2}\right)$ are
axial and vector form factors of $B_{c}\to\psi W$ transition. It
is clear that these functions cannot be determined from perturbative
theory, so some other approach should be used, such as for example
QCD sum rules \cite{Huang:2007kb,Kiselev:1999sc,Kiselev:2000nf,Kiselev:2000pp,Kiselev:2002vz},
different potential quark models  \cite{Kiselev:1992tx,Gershtein:1994jw,Gershtein:1997qy,Colangelo:1999zn,Ivanov:2005fd},
light-front models  \cite{Anisimov:1998xv,Choi:2009ym,Choi:2009ai},
etc. In our paper we will use form-factors sets presented in paper
\cite{Kiselev:2000pp}.

\begin{comment}
https://arxiv.org/pdf/1307.0953.pdf
\end{comment}

These form factors can be parametrized as
\begin{eqnarray*}
F\left(q^{2}\right) & = & \frac{F\left(0\right)}{1-q^{2}/m_{fit}^{2}},
\end{eqnarray*}
where $q^{2}=\left(p_{1}-p_{2}\right)^{2}$ is the transferred momentum
squared and the values of function at zero and maximal agruments are
listed in table \ref{tab:FF}. 

\begin{table}
\begin{centering}
\begin{tabular}{|c||c|c|c|c|c|}
\hline 
\multicolumn{1}{|c|}{\multirow{1}{*}{}} & $F(0)$ & $F\left(q_{max}^{2}=M_{B_{c}}^{2}-M_{J/\psi}^{2}\right)$ &  & $F(0)$ & $F\left(q_{max}^{2}=M_{B_{c}}^{2}-M_{J/\psi}^{2}\right)$\tabularnewline
\hline 
\hline 
$A_{0}$ & 0.60 & 1.6 & $A_{2}$ & 0.69 & 1.4\tabularnewline
\hline 
$A_{1}$ & 0.63 & 1.3 & $V$ & 1.0 & 2.1\tabularnewline
\hline 
\end{tabular}
\par\end{centering}
\caption{Parameters of the $B_{c}\to\psi^{(')}W$ transition form factors\label{tab:FF}}

\end{table}

All information about the final system $\R$ is hidden in the effective
polarization vector $\epsilon^{(\R)}$. Its explicit form and numerical
values depend on the number of final particles, their types and momenta.
The only general issue is that in the limit of isospin conservation
the relation $q^{\mu}\epsilon_{\mu}^{(\R)}$ should hold, to the contribution
of $A_{0}$ form factor vanishes. More detailed investigation requires
some model assumptions on the physics of the underlying processes.
Our previous works it was shown, that the resonance model gives the
results that are in a good agreement with the experiment. In the framework
of this model the amplitude of the process under consideration is
written in the terms of hadronic resonances with suitable quantum
numbers and the form of the amplitudes is chosen accordingly. In the
present paper we will continue to use such an approach to describe
the production of $\R=K+4\pi,$ $KK+3\pi$, and $7\pi$ states. It
turns out that the corresponding processes can be explained in terms
of previously considered and tested on experiment reactions with $KK\pi$,
$3\pi$, $5\pi$ production. For this reason we will first consider
these processes.

\section{Known Decays}

\begin{figure}
\begin{centering}
\includegraphics[width=0.9\paperwidth]{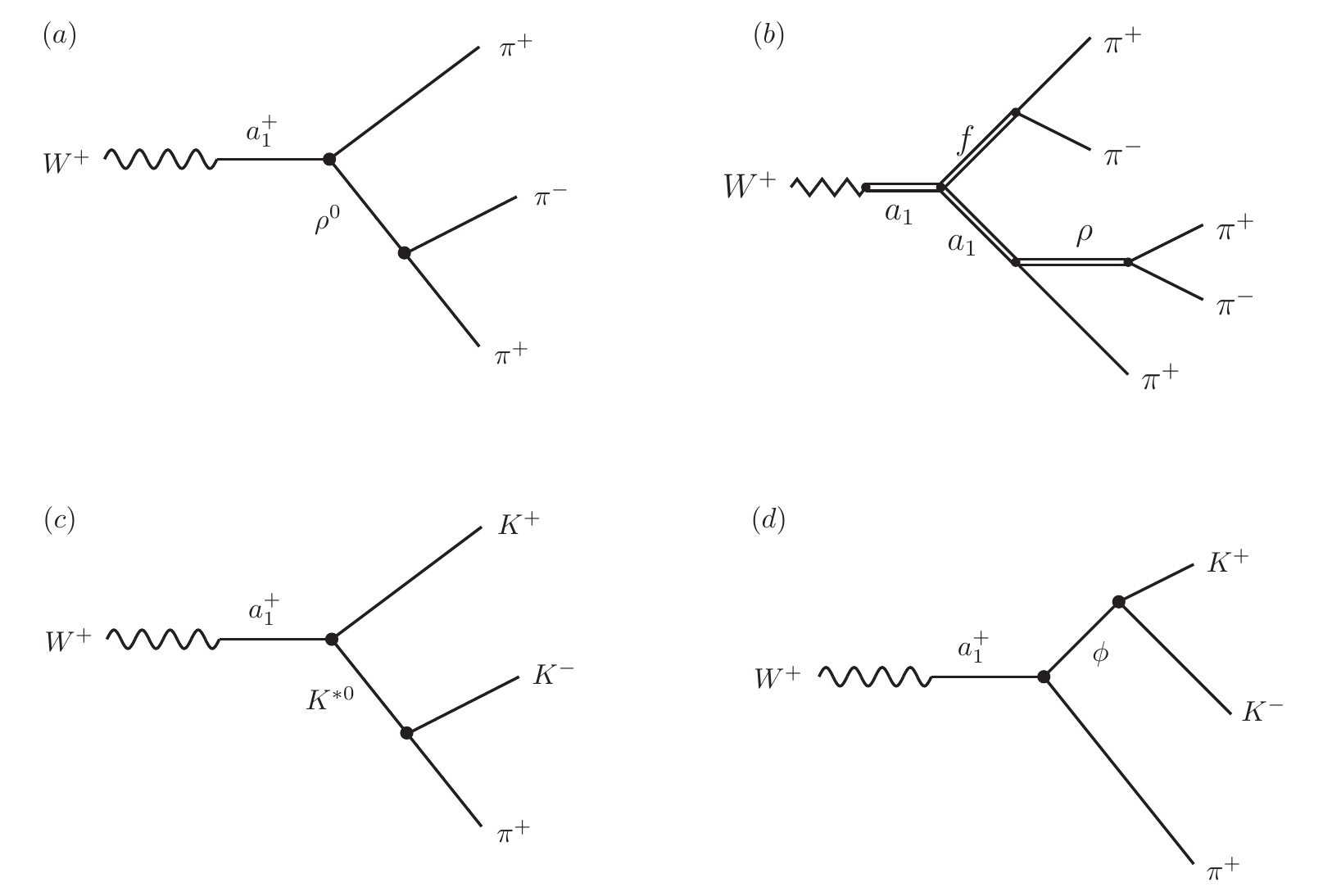}
\par\end{centering}
\caption{Feynman diagrams for old decay processes\label{diags_all_old}}
\end{figure}

\begin{comment}
\begin{figure}
\begin{centering}
\includegraphics[width=0.3\textwidth]{figs/dist_3pi_Q2.pdf}\includegraphics[width=0.3\textwidth]{figs/dist_5pi_Q2.pdf}\includegraphics[width=0.3\textwidth]{figs/dist_KKpi_Q2.pdf}
\par\end{centering}
\caption{Known distributions}
\end{figure}
\end{comment}

\begin{figure}
\begin{centering}
\includegraphics[width=1\textwidth]{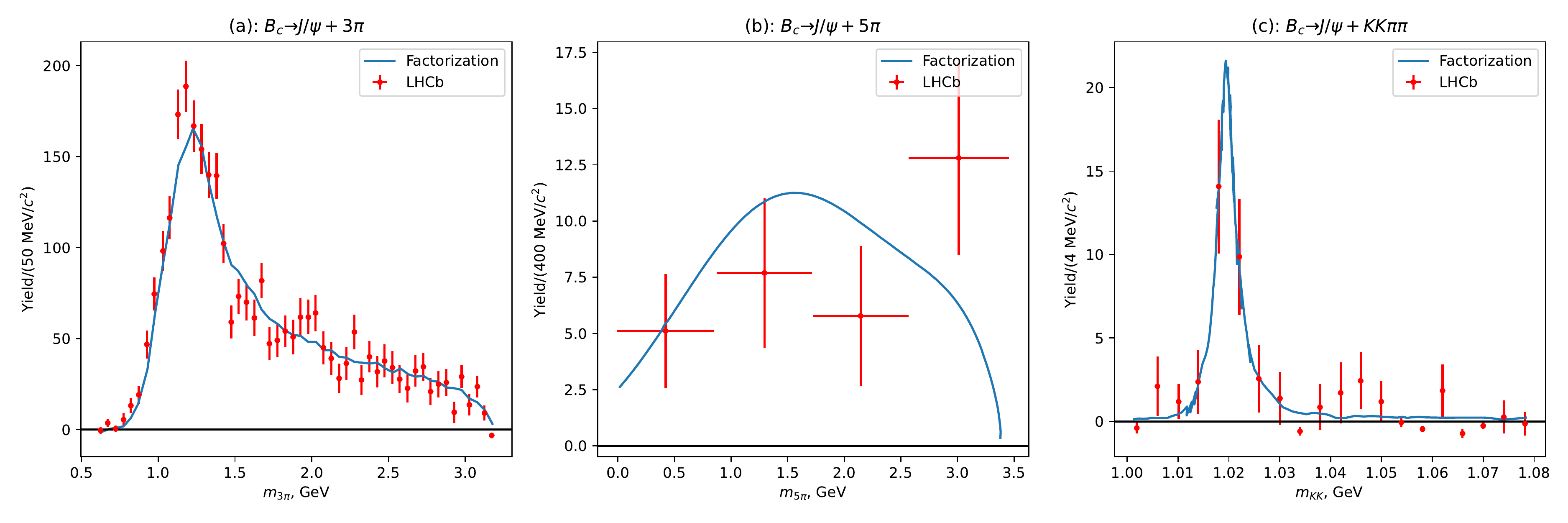}
\par\end{centering}
\caption{Comparison of theoretical predictions (blue line) with experiment
(red dots).LHCb results shown in the left, center and right figures
were published in papers \cite{LHCb:2021tdf}, \cite{LHCb:2014acd}
and \cite{LHCb:2021tdf} respectively.\label{dist_all_old}}

\end{figure}

Let us first consider $B_{c}\to\psi+3\pi$ decay. In resonance approximation
it can be described by the diagram shown on figure \ref{diags_all_old}(a).
The corresponding amplitude can be written in the form
\begin{eqnarray*}
\epsilon_{(3\pi)}^{\mu}\left(Q\to k_{1}k_{2}k_{3}\right) & \sim & BW_{a1}\left(Q^{2}\right)\left(g_{\mu\alpha}-\frac{Q_{\mu}Q_{\alpha}}{Q^{2}}\right)q_{[13]}^{\alpha}BW_{\rho}\left(q_{\{13\}}^{2}\right)+\left\{ q_{1}\Leftrightarrow q_{2}\right\} ,
\end{eqnarray*}
where $q_{1,2,3}$ are the momenta of two final $\pi^{+}$ mesons
and $\pi^{-}$ meson, $Q=q_{1}+q_{2}+q_{3}$ is the total momentum
of virtual $a_{1}$, and notations
\begin{eqnarray*}
q_{[ij]}^{\alpha} & = & q_{i}^{\alpha}-q_{j}^{\alpha},\qquad q_{\{ij\}}^{\alpha}=q_{i}^{\alpha}+q_{j}^{\alpha}
\end{eqnarray*}
were introduced. As you can see, we are using usual Feynman rules
to write this relation and general from of the vertices is determined
by quantum numbers of the participating particles. For example, the
interaction $(q_{1}-q_{3})_{\alpha}$ was selected for $\rho\to\pi\pi$
decay since we are dealing with vector particle decay into two scalars,
so P-wave is in place. The only interesting thing in the above equation
is the propagators of the virtual $a_{1}$ and $\rho$ mesons. Following
\cite{Flatte:1976dk} they are written using Flatte parametrization,
where the energy dependence of the resonance width is taken into account:
\begin{eqnarray*}
BW_{a}\left(q^{2}\right) & = & \frac{M_{a}^{2}}{M_{a}^{2}-q^{2}-iM_{a}\Gamma_{a}(q)},\\
BW_{\rho}\left(q^{2}\right) & = & \frac{1}{1+\beta}\left[\frac{M_{\rho}^{2}}{M_{\rho}^{2}-q^{2}-iM_{\rho}\Gamma_{\rho}\left(q^{2}\right)}+\beta\frac{M_{\rho'}^{2}}{M_{\rho'}^{2}-q^{2}-iM_{\rho'}\Gamma_{\rho'}\left(q^{2}\right)}\right]
\end{eqnarray*}
This process was studied theoretically in works \cite{Berezhnoy:2011nx,Luchinsky:2012rk,Kuhn:2006nw}and
later experimental confirmation by the LHCb collaboration followed
\cite{LHCb:2012ag} As you can see from figure \ref{dist_all_old}(a),
the agreement between theory and experiment is pretty good.

This process can be used as a building block to describe the production
of the higher number of charged $\pi$ mesons, the corresponding Feynman
diagram is shown in figure \ref{diags_all_old}(b). The amplitude
of this diagram is equal to
\begin{eqnarray*}
\epsilon_{(5\pi)}^{\mu} & = & BW_{a}\left(Q^{2}\right)BW_{f}\left(\left(q_{4}+q_{5}\right)^{2}\right)\epsilon_{(3\pi)}^{\mu}\left(q_{1},q_{2},q_{3}\right)+\text{symmetrization},
\end{eqnarray*}
and the comparison of theoretical \cite{Luchinsky:2012rk} and experimental
\cite{LHCb:2016hbr} distributions in shown in figure \ref{dist_all_old}(b).
As you can see, the agreement is again quite reasonable.

Let is consider now production of $KK\pi$ state. Two Feynman diagrams,
that were used to describe it, are shown in figure \ref{diags_all_old}(c),
(d). The corresponding amplitude can be written in the following form
\begin{eqnarray*}
\epsilon_{KK\pi}^{\mu} & = & BW_{a}\left(Q^{2}\right)BW_{K^{*}}\left(q_{\{12\}}^{2}\right)q_{[12]}^{\mu}
\end{eqnarray*}
Figure \ref{dist_all_old}(c) shows that there is a good agreement
between theoretical \cite{Luchinsky:2013yla} and experimental \cite{LHCb:2021tdf}
results.

\section{New Decays}

In this section three new decays will be described, we will present
amplitudes of these processes and theoretical predictions for some
typical distributions.

\begin{figure}
\begin{centering}
\includegraphics[scale=0.9]{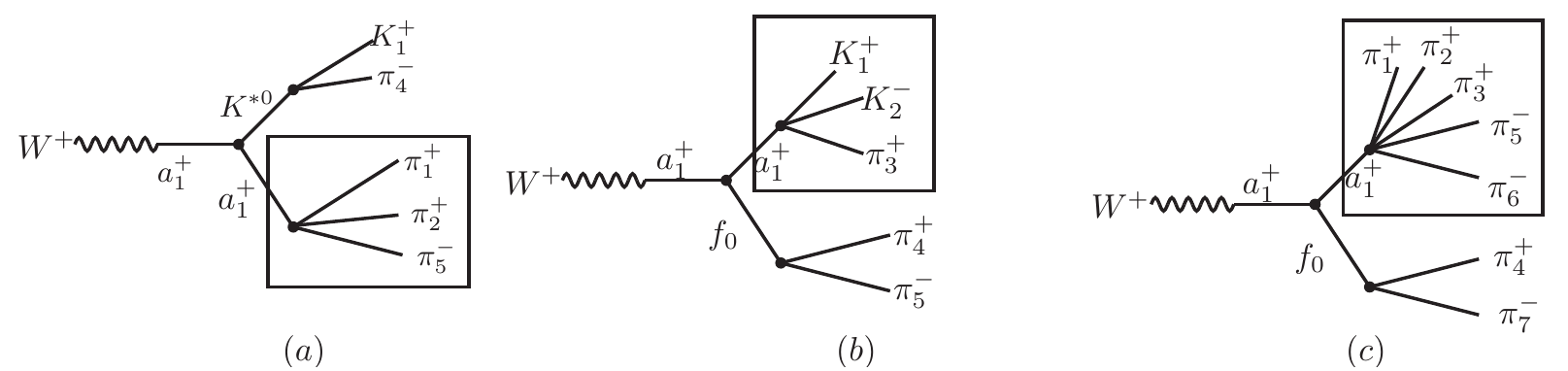}
\par\end{centering}
\caption{Diagram for $W\to\R$ transition\label{diag:all_new}}
\end{figure}

\begin{figure}
\begin{centering}
\includegraphics[width=0.9\textwidth]{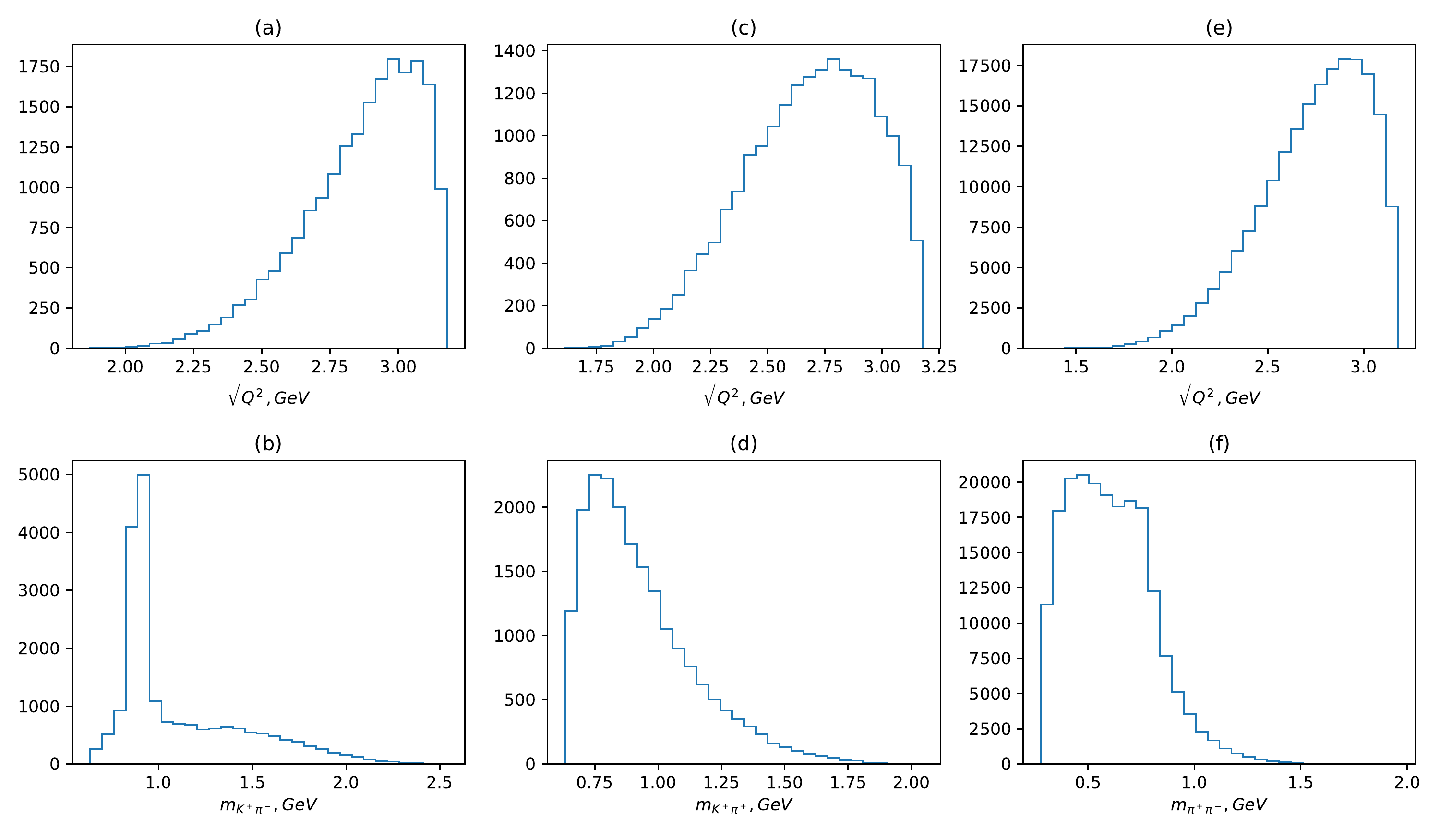}
\par\end{centering}
\caption{Distributions in $B_{c}\to J/\psi+\R$ decays:\label{dist:all_new}}
\end{figure}

Let us consider $\R=K^{+}\pi^{+}\pi^{+}\pi^{-}\pi^{-}$ final state
first. In the resonance approximation $W$-boson hadronization into
this system of light particles can be described by the diagram shown
in figure \ref{diag:all_new}(a). As you can see, this diagram includes
$a_{1}\to3\pi$ transition as a subprocess, so we can use presented
above parametrization to describe it. The form of the other vertices'
amplitudes is easily determined by the quantum numbers of the interacting
particles. As a result, the following form of the transition amplitude
was used later:
\begin{eqnarray*}
\epsilon_{K4\pi}^{\mu} & = & BW_{a}\left(Q^{2}\right)BW_{K^{*}}\left(q_{\{14\}}^{2}\right)e^{\mu\nu\alpha\beta}q_{[14]}^{\nu}\left(q_{\{14\}}-q_{\{234\}}\right)^{\alpha}\epsilon_{(3\pi)}^{\beta}\left(q_{2,}q_{3,}q_{5}\right).
\end{eqnarray*}
In figure \ref{dist:all_new}(a), (b) some distributions are shown.
On the last figure in the $m_{K\pi}$ system you can clearly see a
peak caused by $K^{*0}$ resonance.

The subprocess $a_{1}\to KK\pi$, on the other hand, can be used in
production of $KK+3\pi$ final state. The diagram is shown in figure
\ref{diag:all_new}(b) and the corresponding can be written in the
form
\begin{eqnarray*}
\epsilon_{KK3\pi}^{\mu} & = & BW_{a}\left(Q^{2}\right)BW_{f}\left(q_{\{45\}}^{2}\right)\epsilon_{KK\pi}^{\mu}\left(q_{1}q_{2}q_{3}\right)+\text{symmetrization}.
\end{eqnarray*}
In figure \ref{dist:all_new}(c), (d) some distributions of the corresponding
hadronic process are shown. Note that in this case there are no peaks
in the $m_{K\pi}$ spectrum since there are no resonances in $K^{+}\pi^{+}$
channel.

The final process to be considered in our article is $B_{c}\to\psi^{(')}+7\pi$
decay. The corresponding $W$-boson hadronization diagram is shown
in figure \ref{diag:all_new}(c).. As you can see, this reaction includes
$a_{1}\to5\pi$ transition as a subprocess, to its amplitude is written
as
\begin{eqnarray*}
\epsilon_{7\pi}^{\mu} & = & BW_{a}\left(Q^{2}\right)BW_{f}\left(q_{\{47\}}^{2}\right)\epsilon_{5\pi}^{\mu}\left(q_{1,}q_{2},q_{3},q_{5},q_{6}\right)+\text{symmetrization}.
\end{eqnarray*}
In figure \ref{dist:all_new}(e), (f) distributions over the transferred
momentum and mass of the $\pi^{+}\pi^{-}$ pair are shown. In the
latter distribution you can see a small $\rho$-meson peak, which
is almost hidden by the broad $f_{0}$ resonance.

\section{Conclusion}

The presented paper is devoted to theoretical analysis of some exclusive
$B_{c}$-meson decays. This work is a continuation of a series of
papers, that are using the same approach for description of some other
decays. 

The theoretical model used in all these works is pretty simple. The
reactions under consideration are represented as two-step process.
The first step is weak $B_{c}$-meson decay into final victor charmonium
and virtual $W$-boson, which then hadronizes into a system of light
mesons. The first stage was described in terms of $B_{c}$-meson form-factors,
while for the second stage the resonance approximation was used.

This simple method turned out to be surprisingly useful and able to
produce theoretical predictions for a number of Bc-meson decays, that
are in good agreement with experimental results. In the presented
paper it was used to obtain analytical expressions for the amplitudes
of three more decays: $B_{c}\to J/\psi+K+4\pi$, $B_{c}\to J/\psi+KK+3\pi$,
and $B_{c}\to J/\psi+7\pi$. In addition, the numerical analysis of
these reactions was performed and same interesting differential distributions
can be found in thins work.

There is, of course, lot of work to be done in this field. For example,
there are same unknown normalization constants in the used model,
that cannot be determined from experimental data. For this reason
we do not make predictions for the branching fractions of the considered
decays, only normalized distributions are presented. Calculation of
the branching fractions will be the topic of our future work.

The author would like to thank Dr. A. Likhoded, Dr. I. Belyaev, and Dr. D. Pereima
for useful, stimulating, and motivating discussions. This research
was done with support of RFBR grant 20-02-00154 A.

% \bibliographystyle{apsrev}
% \bibliography{bc_new_litr}

\end{document}